\documentclass[aps,prl,showpacs,twocolumn,amsmath,10pt,superscriptaddress,floatfix,nofootinbib,a4paper]{revtex4-1}
\begin{document}
\title{Establishing the vector-meson-exchange dominance for the short range interactions of light quarks}

\author{Bing-Song Zou}
\email{zoubs@mail.tsinghua.edu.cn}

\affiliation{Department of Physics, Tsinghua University, Beijing 100084, China}\affiliation{CAS Key Laboratory of Theoretical Physics, Institute of Theoretical Physics, Chinese Academy of Sciences, Beijing 100190, China}
\affiliation{School of Physical Sciences, University of Chinese Academy of Sciences, Beijing 100049, China}
\affiliation{Southern Center for Nuclear-Science Theory (SCNT), Institute of Modern Physics, Chinese Academy of Sciences, Huizhou 516000, China}

\begin{abstract}
I give a brief comment on a recent study revealing the vector meson exchange (VME) dominant for the short range interactions between u/d quarks in the $NN$, $D_{03}$, and $D_{30}$ systems. The finding echoes nicely with an earlier study of hadron spectroscopy using a quark model with hidden local symmetry which also favors the VME dominant for the short range interactions of light quarks. The VME dominance for the short range interactions of
light quarks gives a natural explanation of the empiric
VME dominance for the interactions between hadrons containing
light quarks.
\end{abstract}

\maketitle

Understanding the interactions between constituent quarks is
one of the fundamental and challenging problems in hadron
physics. At present, the constituent quark models have
achieved great success in describing the properties of hadrons,
whereas some long-standing disputes remain, especially
for the short range interactions between constituent
quarks. In exploring the dynamic mechanism between u/d
quarks, it is found that the one gluon exchange and/or vector
meson exchange induced interactions may dominate the
short range interactions~\cite{Dai:2003dz}. However, owing to the complexity
of the strong interaction, it remains unsettled which
one dominates the short range interactions between u/d
quarks, one gluon exchange, vector meson exchange, or both
of them. There is no doubt that the resolution of this problem
will uncover the mechanism of short range interactions and
deepen our understanding of the strong interaction in the low
energy QCD region.
Recently, L\"u et al.~\cite{Lu:2024dtb} dedicated themselves to revealing
the short range interactions between u/d quarks by systematically
studying the properties of non-strange dibaryons. For
the first time, the $NN$, $D_{03}$, and $D_{30}$ systems were jointly
investigated in the extended chiral SU(3) quark models, and
the comparisons of these three distinct systems offer a novel
and practical way to explore the short range dynamic mechanism.
In the chiral SU(3) quark model, the potentials
include the one gluon exchange terms, confinement, and the
interactions induced by the scalar and pseudo-scalar nonet
chiral fields. Meanwhile, the vector meson fields are introduced
in the extended chiral SU(3) quark model, which is
more suitable to investigate the relative strengths between
one gluon exchange and vector meson exchange interactions.
For the properties of the $NN$ system, the extended chiral
SU(3) quark model with different kinds of coupling
strengthens can describe the experimental data well. However,
the deuteron is a loosely bound state of two nucleons
interacting with each other primarily through the long range
pion meson exchange, insensitive to the short range interactions.
Therefore, the ideal systems are the compact dibaryon
$D_{03}$ as well as its mirror state $D_{30}$. It is found that to
explain simultaneously the available data in the $NN$, $D_{03}$, and
$D_{30}$ systems~\cite{Clement:2016vnl}, a relatively large contribution from the
vector meson exchange (VME) is necessary while a small
residual one gluon exchange interaction is allowed. With the
extended chiral quark model of VME dominance, the binding
energies of $D_{03}$ and $D_{30}$ as $\Delta\Delta$ bound states are 58-111
MeV and 5-12 MeV, respectively, corresponding to the experimentally
observed deeply $D_{03}$ bound state $d^*(2380)$ and a
possible weakly $D_{30}$ bound state~\cite{Clement:2016vnl} perfectly. Note that
earlier symmetry analysis without VME expected similar
binding energies for the $D_{03}$ and $D_{30}$ systems.

The strong indication of the VME dominance for the short
range interactions between light quarks from the study of
dibaryons~\cite{Lu:2024dtb} echoes nicely with a recent study of hadron
spectroscopy using a quark model with hidden local symmetry~\cite{He:2023ucd}. There, vector mesons are included based on local
hidden symmetry and play an important role for reproducing
meson and baryon spectra simultaneously, mainly owing to
the effects of $\omega$ meson exchange—attractive for antiquarkquark
and repulsive for quark-quark. In addition, the predicted
mass of $T_{cc}$ tetraquark state is much closer to its experimental
value than the result without including the VME.
The VME dominance for the short range interactions of
light quarks also gives a natural explanation of the empiric
VME dominance for the interactions between hadrons containing
light quarks~\cite{Dong:2021bvy}. Based on VME dominance for the
hadron interactions, many hadronic molecules composed of a
pair of charmed hadrons have been predicted, including
doubly charmed dibaryons~\cite{Dong:2021bvy}. Finding these predicted
doubly charmed hadronic molecules would be extremely
important for establishing the VME dominance for the short
range interactions of light quarks. Of course, future theoretical
and experimental explorations on various other dibaryons
or other kinds of hadronic molecules\cite{Liu:2022gxf,Chen:2024eaq} would also
be very useful to pin down the dynamic mechanism of short
range interactions.

%\bigskip
%\noindent {\bf Acknowledgments:}
%This work is supported by the NSFC and the Deutsche Forschungsgemeinschaft (DFG, German Research
%Foundation) through the funds provided to the Sino-German Collaborative
%Research Center TRR110 “Symmetries and the Emergence of Structure in QCD”
%(NSFC Grant No. 12070131001, DFG Project-ID 196253076 - TRR 110), by the NSFC 
%Grant No.11835015, No.12047503, and by the Chinese Academy of Sciences (CAS) under Grant No.XDB34030000.



\begin{thebibliography}{}
%Journal Author, Journal \textbf{Volume}, page numbers (year)
%Book Author, \textit{Book title} (Publisher, place, year) page numbers
% etc
%

%\cite{Dai:2003dz}
\bibitem{Dai:2003dz}
L.~R.~Dai, Z.~Y.~Zhang, Y.~W.~Yu and P.~Wang,
%``N N interactions in the extended chiral SU(3) quark model,''
Nucl. Phys. A \textbf{727}, 321-332 (2003)
doi:10.1016/j.nuclphysa.2003.08.006
[arXiv:nucl-th/0404004 [nucl-th]].
%90 citations counted in INSPIRE as of 06 Feb 2025

%\cite{Lu:2024dtb}
\bibitem{Lu:2024dtb}
Q.~F.~L\"u, Y.~B.~Dong, P.~N.~Shen and Z.~Y.~Zhang,
%``Reveal short range interactions between u/d quarks in the NN, D$_{03}$, and D$_{30}$ systems,''
Sci. China Phys. Mech. Astron. \textbf{68}, no.3, 232011 (2025)
doi:10.1007/s11433-024-2541-9
[arXiv:2407.01993 [hep-ph]].
%0 citations counted in INSPIRE as of 06 Feb 2025

%\cite{Clement:2016vnl}
\bibitem{Clement:2016vnl}
H.~Clement,
%``On the History of Dibaryons and their Final Observation,''
Prog. Part. Nucl. Phys. \textbf{93}, 195 (2017)
doi:10.1016/j.ppnp.2016.12.004
[arXiv:1610.05591 [nucl-ex]].
%128 citations counted in INSPIRE as of 06 Feb 2025

%\cite{He:2023ucd}
\bibitem{He:2023ucd}
B.~R.~He, M.~Harada and B.~S.~Zou,
%``Quark model with hidden local symmetry and its application to Tcc,''
Phys. Rev. D \textbf{108}, no.5, 054025 (2023)
doi:10.1103/PhysRevD.108.054025
[arXiv:2306.03526 [hep-ph]].
%10 citations counted in INSPIRE as of 06 Feb 2025

%\cite{Dong:2021bvy}
\bibitem{Dong:2021bvy}
X.~K.~Dong, F.~K.~Guo and B.~S.~Zou,
%``A survey of heavy\textendash{}heavy hadronic molecules,''
Commun. Theor. Phys. \textbf{73}, no.12, 125201 (2021)
doi:10.1088/1572-9494/ac27a2
[arXiv:2108.02673 [hep-ph]].
%181 citations counted in INSPIRE as of 06 Feb 2025

%\cite{Liu:2022gxf}
\bibitem{Liu:2022gxf}
H.~Liu, J.~He, L.~Liu, P.~Sun, W.~Wang, Y.~B.~Yang and Q.~A.~Zhang,
%``Exploring hidden-charm and hidden-strange hexaquark states from lattice QCD,''
Sci. China Phys. Mech. Astron. \textbf{67}, no.1, 211011 (2024)
doi:10.1007/s11433-023-2205-0
[arXiv:2207.00183 [hep-lat]].
%13 citations counted in INSPIRE as of 06 Feb 2025

%\cite{Chen:2024eaq}
\bibitem{Chen:2024eaq}
J.~Chen, F.~K.~Guo, Y.~G.~Ma, C.~P.~Shen, Q.~Shou, Q.~Wang, J.~J.~Wu and B.~S.~Zou,
%``Production of exotic hadrons in $pp$ and nuclear collisions,''
doi:10.1007/s41365-025-01664-w
[arXiv:2411.18257 [hep-ph]].
%1 citations counted in INSPIRE as of 06 Feb 2025
 
\end{thebibliography}
\end{document}